\documentclass{article}
\usepackage{makeidx}
\usepackage{amsmath}
\usepackage{amsfonts}
\usepackage{amssymb}
\usepackage{epsfig}
\usepackage{psfrag}

\newtheorem{thm}{Theorem}

\newtheorem{prop}[thm]{Proposition}

\def\bra #1{\langle #1\vert}
\def\ket #1{\vert #1\rangle}
\def\braket #1#2{\langle #1 \vert #2\rangle}

\begin{document}

\title{Gaussian maximizers for quantum Gaussian observables and ensembles}
\author{A. S. Holevo \\
Steklov Mathematical Institute \\
Gubkina 8, 119991 Moscow, Russia}
\date{}
\maketitle

\begin{abstract}
In this paper we prove two results related to the Gaussian optimizers
conjecture for multimode bosonic system with gauge symmetry. First, we argue
that the classical capacity of a Gaussian observable is attained
on a Gaussian ensemble of coherent states. This generalizes results
previously known for heterodyne measurement in one mode. By using this fact
and continuous variable version of ensemble-observable duality, we prove an
old conjecture that accessible information of a Gaussian ensemble is
attained on the multimode generalization of the heterodyne measurement.
\end{abstract}

\section{Introduction}

In this paper we prove two results related to the Gaussian optimizers
conjecture for multimode bosonic system with global gauge symmetry\footnote{%
In quantum communication literature such systems are called \emph{%
phase-insensitive}.}. In theorem \ref{t1} of sec. \ref{s2} we argue that the
classical capacity of an arbitrary gauge-covariant Gaussian observable --
considered as a communication channel with quantum input and classical
output -- is attained on a Gaussian ensemble of coherent states. This
generalizes result previously known for the heterodyne measurement \cite%
{hall}. In the difficult part of the argument -- the minimization of the
output differential entropy -- we rely upon our previous result \cite{h4}
obtained as a limiting case of the general solution of the Gaussian
optimizers conjecture for quantum Gaussian channels \cite{ghg}. Let us
stress that it is not possible to apply that solution directly to a Gaussian
observable because there is no way to embed a continuously-valued observable
(as distinct from discretely-valued observables) into a channel with quantum
output \cite{h5}. The classical capacity of observable is the most important
quantity characterizing the ultimate information-processing performance of
the measurement (see e.g. \cite{hall}, \cite{da}, \cite{h5}).

By using theorem \ref{t1} and infinite-dimensional version of
ensemble-observable duality developed in sec. \ref{s3}, we prove the main
result of this work -- theorem \ref{t2} concerning accessible information of
a Gaussian ensemble. In particular, it answers an old conjecture \cite{h6},
\cite{b1}, \cite{b2} that the accessible information of a Gaussian ensemble
is attained by the multimode generalization of the heterodyne measurement.
As in the other Gaussian optimizer problems, the difficulty here lies in
finding the \textit{global maximum } of a \textit{convex} functional, when
the optimal solution turns out to be highly non-unique and the standard
tools of convex analysis become inefficient.

\section{Preliminaries}

Let $\mathcal{M}=\{M(dx)\}$ be an observable (POVM) in a separable Hilbert
space $\mathcal{H}$ with the outcome space $\mathcal{X}$ \ which is a
complete separable metric space. A corresponding measurement channel is
defined as transformation $\mathcal{M}:\rho \longrightarrow \mathrm{Tr}\rho
M(dx)$ of density operators (d.o.) $\rho $ to probability distributions on $%
\mathcal{X}$. In \cite{h1} the existence of a $\sigma -$finite measure $\mu
(dx)$ was shown such that for any d.o. $\rho $ the probability measure $%
\mathrm{Tr}\rho M(dx)$ is absolutely continuous w.r.t. $\mu (dx),$ thus
having the probability density (p.d.) $p_{\rho }(x).$ Therefore the \textit{%
\ measurement channel} can be defined as the transformation
\begin{equation*}
\mathcal{M}:\rho \rightarrow p_{\rho }
\end{equation*}%
mapping affinely d.o. on $\mathcal{H}$ into p.d. on $\left( \mathcal{X},\mu
\right) $. Notice that $\mu (dx)$ is defined uniquely only up to the class
of mutually absolutely continuous measures.

A (generalized) \textit{ensemble} $\mathcal{E}=\left\{ \pi (dx),\rho
_{x}\right\} $ consists of probability measure $\pi (dx)$ on the input space
$\mathcal{X}$ and a measurable family of d.o. $\rho _{x}$ on $\mathcal{H}$.
The \textit{average state} of the ensemble is the barycenter of this measure%
\begin{equation*}
\bar{\rho}_{\mathcal{E}}=\int_{\mathcal{X}}\rho _{x}\,\pi (dx),
\end{equation*}%
the integral existing in the strong sense in the Banach space of trace class
operators. Let $\mathcal{M}=\{M(dy)\}$ be an observable with the outcome
space $\mathcal{Y}$ \ and $\rho \rightarrow p_{\rho }$ the corresponding
measurement channel. The joint probability distribution of $x,y$ on $%
\mathcal{X\times Y}$ \ is uniquely defined by the relation%
\begin{equation*}
P(A\times B)=\int_{A}\pi (dx)\mathrm{Tr}\,\rho _{x}M(B)=\mathrm{Tr}%
\int_{A}\int_{B}\,p_{\rho _{x}}(y)\,\pi (dx)\mu (dy),
\end{equation*}%
where $A$ is an arbitrary Borel subset of $\mathcal{X}$ and $B$ is that of $%
\mathcal{Y}.$

The classical Shannon information between $x,y$ is equal to (cf. \cite{dobr})%
\begin{equation*}
I(\mathcal{E},\mathcal{M})=\int \int \pi (dx)\mu (dy)p_{\rho _{x}}(y)\log
\frac{p_{\rho _{x}}(y)}{p_{\bar{\rho}_{\mathcal{E}}}(y)}.
\end{equation*}%
We will use the differential entropy
\begin{equation*}
h(p)=-\int p(x)\log p(x)\mu (dx)
\end{equation*}%
of a p.d. $p(x).$ There is a special class $\mathcal{D}$ of p.d.'s we will
be using for which the differential entropy is well-defined. Let $\mathcal{X}
$ \ be a $d$-dimensional vector space and $p(x)$ a bounded p.d. on $\mathcal{%
X}$ such that $p(x)\leq c^{d}$ (mod $\mu $) for some $c>0$. Then $h(p)$ is
well-defined with values in $[d\log c,+\infty ]$ because in this case $%
\tilde{p}(x)=p(x/c)c^{-d}$ is a p.d. satisfying $\tilde{p}(x)\leq 1$ (mod $%
\mu $), hence $-\tilde{p}(x)\,\log \,\tilde{p}(x)\,\geq 0$. Thus $h(\tilde{p}%
)$ is well-defined with values in $[0,+\infty ]$ and by change of variable $%
\tilde{x}=cx$,
\begin{equation}
h(p)=h(\tilde{p})+d\log c  \label{K}
\end{equation}%
is also well-defined with values in $[d\log c,+\infty ]$.

If observable $\mathcal{M}$ is such that $p_{\rho }\in \mathcal{D}$ for any
d.o. $\rho$ and ensemble $\mathcal{E}$ is such that $h(p_{\bar{\rho}_{%
\mathcal{E}}})<\infty$, then the Shannon information between $x,y$ is equal
to%
\begin{equation}
I(\mathcal{E},\mathcal{M})=h(p_{\bar{\rho}_{\mathcal{E}}})-\int h(p_{\rho
_{x}})\pi (dx).  \label{rchi}
\end{equation}%
This quantity is well-defined with values in $[0,+\infty ]$ due to Jensen's
inequality.

In what follows $\mathcal{H}$ will be the space of a strongly continuous
irreducible projective unitary representation $z\rightarrow D(z)$ of the
canonical commutation relations (CCR) (see e.g. \cite{asp}, \cite{h} for a
detailed account) describing quantization of a linear classical system with $%
s$ degrees of freedom such as finite number of physically relevant
electromagnetic modes in a receiver's cavity.

The classical linear system with the preferred complex structure (gauge) is
described by the phase space $\mathbb{C}^{s}$ equipped with the symplectic
form $2\mathrm{Im\,\ }z^{\ast }w,$ where
\begin{equation*}
z=\left[
\begin{array}{c}
z_{1} \\
\dots \\
z_{s}%
\end{array}%
\right] ,\quad z^{\ast }=\left[
\begin{array}{ccc}
\bar{z}_{1} & \dots & \bar{z}_{s}%
\end{array}%
\right] .
\end{equation*}%
We will use the symplectic Fourier transform%
\begin{equation*}
\tilde{f}(w)=\int \exp \left( z^{\ast }w-w^{\ast }z\right) \,f(z)\frac{%
d^{2s}z}{\pi ^{s}}=\int \exp \left( 2i\mathrm{Im\,}z^{\ast }w\right) f(z)%
\frac{d^{2s}z}{\pi ^{s}}.
\end{equation*}%
Notice that $\tilde{\tilde{f}}=f$ i.e. inverse transform has the same form.

The quantization gives a bosonic system described by the collection of
annihilation-creation operators, in the vector form%
\begin{equation*}
a=\left[
\begin{array}{c}
a_{1} \\
\dots \\
a_{s}%
\end{array}%
\right] ,\quad a^{\dagger }=\left[
\begin{array}{ccc}
a_{1}^{\dagger } & \dots & a_{s}^{\dagger }%
\end{array}%
\right] ,
\end{equation*}%
where the lower index of a component refers to the number of the mode. The
CCR including the nonvanishing commutator%
\begin{equation*}
a_{j}a_{k}^{\dagger }-a_{k}^{\dagger }a_{j}=\delta _{jk}I,
\end{equation*}%
are conveniently written in terms of displacement operators $D(z)=\exp
\left( a^{\dagger }z-z^{\ast }a\right) ,$ namely
\begin{equation}
D(z)D(w)=\exp \left( -i\,\mathrm{Im\,}z^{\ast }w\right) D(z+w),\quad z,w\in
\mathbb{C}^{s}.  \label{ccr}
\end{equation}

The (global) gauge group acts as $z\rightarrow e^{i\varphi }z,$ ($\varphi $
is real phase) in the classical space, and via the unitary group $\varphi
\rightarrow U_{\varphi }=\exp \left( -i\varphi \,a^{\dagger }a\right) $ in $%
\mathcal{H}$ ($a^{\dagger }a$ is the total number operator), so that%
\begin{equation}
U_{\varphi }^{\ast }\,a_{j}U_{\varphi }=a_{j}e^{-i\varphi },\quad U_{\varphi
}^{\ast }\,a_{j}^{\dagger }U_{\varphi }=a_{j}^{\dagger }e^{i\varphi }.
\label{phase}
\end{equation}

The quantum Fourier transform of a trace class operator $\rho $ is defined
as
\begin{equation*}
\mathop{\rm Tr}\nolimits\rho D(w)
\end{equation*}%
The quantum Parceval formula holds:%
\begin{equation}
\mathop{\rm Tr}\nolimits\rho \sigma ^{\ast }=\int \mathop{\rm Tr}%
\nolimits\rho D(w)\,\overline{\mathop{\rm Tr}\nolimits\sigma D(w)}\frac{%
d^{2s}w}{\pi ^{s}}  \label{parc}
\end{equation}

An operator $\rho $ is gauge-invariant if $U_{\varphi }^{\ast }\,\rho
U_{\varphi }=\rho $ for all values of the phase $\varphi$. A gauge-invariant
Gaussian d.o. is defined by the quantum characteristic function\footnote{%
We denote $I_{s}$ the unit $s\times s$-matrix, as distinct from the unit
operator $I$ in a Hilbert space.}
\begin{equation}
\mathop{\rm Tr}\nolimits\rho _{\Lambda }D(w)=\exp \left[ -w^{\ast }\left(
\Lambda +\frac{I_{s}}{2}\right) w\right] ,  \label{one-mode-cf}
\end{equation}%
where $\Lambda =\mathrm{Tr}\,a\rho _{\Lambda }a^{\dagger }$ is the complex
covariance matrix satisfying $\Lambda\geq 0$. Notice that $\rho _{\Lambda
}^{\top }=\rho _{\bar{\Lambda}},$ where $^{\top }$ denotes the transposition
operation defined in (\ref{trans}) (see Appendix A).

In the Hilbert space $\mathcal{H}$ of an irreducible representation of CCR
there is a unique unit vacuum vector $|0\rangle $ such that $a|0\rangle =0.$
The case $\Lambda =0$ in (\ref{one-mode-cf}) corresponds to the vacuum d.o. $%
\rho _{0}=|0\rangle \langle 0|.$ The coherent state vectors are $|z\rangle
=D(z)|0\rangle .$

We will use the P-representation in the case of nondegenerate $\Lambda :$
\begin{equation}
\rho _{\Lambda }=\int |z\rangle \langle z|\exp \left( -z^{\ast }\Lambda
^{-1}z\right) \frac{d^{2s}z}{\pi ^{s}\det \Lambda },  \label{noshifted}
\end{equation}%
Another important Gaussian d.o. is obtained by action of the displacement
operators
\begin{equation}
\rho _{\Lambda ,z}=D(z)\rho _{\Lambda }D(z)^{\dagger }=\int |w\rangle
\langle w|\exp \left( -\left( w-z\right) ^{\ast }\Lambda ^{-1}\left(
w-z\right) \right) \frac{d^{2s}w}{\pi ^{s}\det \Lambda },  \label{shifted2}
\end{equation}%
it has the quantum characteristic function
\begin{equation}
\mathop{\rm Tr}\nolimits\rho _{\Lambda ,z}D(w)=\exp \left[ 2i\mathrm{Im\,}%
z^{\ast }w-w^{\ast }\left( \Lambda +\frac{I_{s}}{2}\right) w\right] .
\label{lamz}
\end{equation}

\section{Gaussian observables and ensembles}

\label{s2}

In this Section we will consider Gaussian observables with the outcome space
$\mathbb{C}^{s},$ described by POVM
\begin{equation}
\tilde{M}(d^{2s}z)=D(Kz)\rho _{N}D(Kz)^{\dagger }\frac{\left\vert \det
K\right\vert ^{2}d^{2s}z}{\pi ^{s}};\quad z\in \mathbb{C}^{s},  \label{MT}
\end{equation}%
where $K$ a nondegenerate complex $s\times s-$matrix, and d.o. $\rho _{N}$
is defined by (\ref{one-mode-cf}) with $\Lambda =N.$ This is a special
(gauge-covariant) case of general Gaussian observables considered in \cite{h}%
. Particularly important is the case $K=I_{s},$ where
\begin{equation}
M(d^{2s}z)=D(z)\rho _{N}D(z)^{\dagger }\frac{d^{2s}z}{\pi ^{s}};\quad z\in
\mathbb{C}^{s}.  \label{M}
\end{equation}%
In the Appendix A we recall alternative description of such observables via
extension to a spectral measure in a composite system including ancillary
system (going back to \cite{h6}). By taking $N=0$ so that $\rho
_{0}=|0\rangle \langle 0|$ is the vacuum state, we obtain the multimode
version of the \textquotedblleft heterodyne measurement\textquotedblright\
\begin{equation}
M_{\ast }(d^{2s}z)=D(z)\rho _{0}D(z)^{\dagger }\frac{d^{2s}z}{\pi ^{s}}%
=|z\rangle \langle z|\frac{d^{2s}z}{\pi ^{s}},  \label{M*}
\end{equation}%
see \cite{ysh}. Thus the POVM (\ref{M}) corresponds to a noisy
generalization of the multimode heterodyne measurement.

Let $\rho $ be an input d.o. then by using (\ref{parc}), (\ref{lamz}) and
real-valuedness of the quadratic form under the exponent, the output p.d. of
the observable (\ref{M}) is
\begin{eqnarray}
p_{\rho }(z) &=&\mathrm{Tr}\,\rho D(z)\rho _{N}D(z)^{\dagger }  \label{pro}
\\
&=&\int \mathrm{Tr}\,\rho D(w)\exp \left[ -2i\mathrm{Im\,}z^{\ast }w-w^{\ast
}\left( N+\frac{I_{s}}{2}\right) w\right] \frac{d^{2s}w}{\pi ^{s}}  \notag
\end{eqnarray}%
and that of observable (\ref{MT}) is $\tilde{p}_{\rho }(z)=p_{\rho
}(Kz)\left\vert \det K\right\vert ^{2}.$ Notice that all these p.d.'s belong
to the class $\mathcal{D}$ because $0\leq \mathrm{Tr}\,\rho \sigma \leq 1$
for any two d.o. $\rho ,\sigma $. Thus the differential entropy of the
output p.d. is well-defined and
\begin{equation}
h(\tilde{p}_{\rho })=h(p_{\rho })-2\log \,\left\vert \det K\right\vert .
\label{sca}
\end{equation}

Let $\Sigma $ be a nonnegative definite complex Hermitian $s\times s-$%
matrix. By $\mathfrak{S}(\Sigma )$ we denote the set of all d.o. with the
\textit{complex covariance matrix}
\begin{equation}
\mathrm{Tr}\,a\rho a^{\dagger }=\Sigma ,  \label{SSigma}
\end{equation}%
There is a unique gauge-invariant Gaussian d.o. $\rho _{\Sigma }$ in $%
\mathfrak{S}(\Sigma )$. By using (\ref{one-mode-cf}) with $\Lambda =\Sigma $
and (\ref{pro}), one obtains the output p.d.
\begin{equation}
\mathrm{Tr}\,\rho _{\Sigma }\,D(z)\,\rho _{N}\,D(z)^{\dagger }=\frac{1}{\pi
^{s}\det \left( \Sigma +N+I_{s}\right) }\exp \left[ -z^{\ast }\left( \Sigma
+N+I_{s}\right) ^{-1}z\right] ,  \label{sipd}
\end{equation}%
which is complex Gaussian p.d. with the covariance matrix $\Sigma +N+I_{s}=%
\tilde{\Sigma}+I_{s}.$

We have%
\begin{equation}
h(p_{\rho })\leq h(p_{\rho _{\Sigma }}),\quad \rho \in \mathfrak{S}(\Sigma ).
\label{mep}
\end{equation}%
Indeed, for any $\rho \in \mathfrak{S}(\Sigma )$ define the gauge-invariant
d.o.
\begin{equation*}
\rho _{gi}=\int\limits_{0}^{2\pi }U_{\varphi }^{\ast }\rho U_{\varphi }\frac{%
d\varphi }{2\pi }.
\end{equation*}%
By the concavity of the differential entropy and Jensen's inequality $%
h(p_{\rho })\leq h(p_{\rho _{gi}}).$ By using (\ref{phase}) it is not
difficult to check that $\rho _{gi}\ $has zero first moments, finite second
moments $\mathrm{Tr}\,a_{j}^{\dagger }\rho a_{k}$ given by (\ref{SSigma})
and other second moments such as $\mathrm{Tr}\,a_{j}\rho a_{k},\,\mathrm{Tr\,%
}a_{j}^{\dagger }\rho a_{k}^{\dagger }$ vanishing. Thus it has all the first
and the second moments the same as the Gaussian d.o. $\rho _{\Sigma }.$ By
the classical maximum entropy principle \cite{cover}, we have $h(p_{\rho
_{gi}})\leq h(p_{\Sigma })$, which proves (\ref{mep}).

We will be interested in the following constrained $\chi -$ capacity of the
channel $\mathcal{M}$%
\begin{equation}
C_{\chi }(\mathcal{M};\Sigma )=\sup_{\mathcal{E}:\bar{\rho}_{\mathcal{E}}\in
\mathfrak{S}(\Sigma )}I(\mathcal{E}, \mathcal{M}),  \label{cchi}
\end{equation}%
where $I(\mathcal{E}, \mathcal{M})$ is the classical information quantity
defined in (\ref{rchi}).

\begin{thm}
\label{t1} Let $\widetilde{\mathcal{M}}$ be the measurement channel
corresponding to the Gaussian observable (\ref{MT}), then the supremum in (%
\ref{cchi}) is equal to
\begin{equation}
C_{\chi }(\widetilde{\mathcal{M}};\Sigma )=\log \det \left( I_{s}+\left(
N+I_{s}\right) ^{-1}\Sigma \right)  \label{cchig}
\end{equation}%
and is attained on the Gaussian ensemble of coherent states
\begin{equation}
\left\{ \exp \left( -z^{\ast }\Sigma ^{-1}z\right) \frac{d^{2s}z}{\pi
^{s}\det \Sigma },\,|z\rangle \langle z|\right\} .  \label{ET}
\end{equation}
\end{thm}

The relation (\ref{cchig}) can be considered as a multimode version of a
formula obtained in \cite{hall} by \textquotedblleft information
exclusion\textquotedblright\ argument.

\textit{Proof}. The channel $\mathcal{M}$ defined by (\ref{M}) is covariant
with respect to the irreducible action of the displacement operators $D(z)$
which means%
\begin{equation*}
D(w)^{\dagger }M(B)D(w)=M(B-w)
\end{equation*}%
for any Borel subset $B\subseteq \mathbb{C}^{s}$ or, equivalently,
\begin{equation*}
p_{D(w)\rho D(w)^{\dagger }}(z)=p_{\rho }(z-w).
\end{equation*}%
By adapting the argument from \cite{h2} for irreducibly covariant quantum
channel to our case of quantum-classical channel, we obtain
\begin{equation}
C_{\chi }(\mathcal{M};\Sigma )=\max_{\rho \in \mathfrak{S}(\Sigma
)}h(p_{\rho })-\check{h}(\mathcal{M}),  \label{anja}
\end{equation}%
where
\begin{equation}
\check{h}(\mathcal{M})=\min_{\rho }h(p_{\rho })  \label{mien}
\end{equation}%
is the minimal output differential entropy. Indeed, from (\ref{rchi}) it
follows that the right-hand side (with `sup' in place of `max' and `inf' in
place of `min') is an upper bound for $C_{\chi }(\mathcal{M};\Sigma )$. Its
achievability follows from Proposition 2 of the recent paper \cite{hk},
since all the assumptions of that result are fulfilled in the Gaussian
gauge-invariant case under consideration here.

To be explicit, first, by a proof of generalization of the Wehrl conjecture
to the measurements of the form (\ref{M}) obtained in \cite{h4}, the minimum
(\ref{mien}) is attained on the vacuum state $\rho _{0}=|0\rangle \langle 0|$
to which corresponds the output p.d.%
\begin{equation*}
p_{\rho _{0}}(z)=\det (N+I_{s})^{-1}\exp \left( -z^{\ast
}(N+I_{s})^{-1}z\right) .
\end{equation*}%
so that
\begin{equation*}
\check{h}(\mathcal{M})=h(p_{\rho _{0}})=\log e^{s}\det (N+I_{s}).
\end{equation*}%
Second, take an ensemble $\mathcal{E}$ such that $\bar{\rho}_{\mathcal{E}%
}\in \mathfrak{S}(\Sigma ).$ Then $p_{\bar{\rho}_{\mathcal{E}}}$ has the
covariance matrix $\Sigma +N+I_{s}$ (see Appendix A). By the maximum entropy
principle (\ref{mep}), the maximum of $h(p_{\bar{\rho}_{\mathcal{E}}})$ is
attained on the Gaussian d.o. $\rho _{\Sigma }$ and is equal to
\begin{equation}
\max_{\rho \in \mathfrak{S}(\Sigma )}h(p_{\rho })=h(p_{\rho _{\Sigma
}})=\log e^{s}\det \left( \Sigma +N+I_{s}\right) .  \label{maxh}
\end{equation}%
And finally, by (\ref{noshifted}) the d.o. $\rho _{\Sigma }$ is the average
state of the Gaussian ensemble $\mathcal{E}$ of coherent states obtained
from the vacuum state $\rho _{0}=|0\rangle \langle 0|$ by the action of the
displacement operators $D(z)$, which is thus the optimal ensemble achieving
the upper bound for (\ref{cchi}).

Thus we obtain the value
\begin{eqnarray}
C_{\chi }(\mathcal{M};\Sigma ) &=&\log \det \left( \Sigma +N+I_{s}\right)
-\log \det \left( N+I_{s}\right)  \notag \\
&=&\log \det \left( I_{s}+\left( N+I_{s}\right) ^{-1}\Sigma \right) .
\label{CX}
\end{eqnarray}%
By taking into account (\ref{sca}) we also obtain that
\begin{equation*}
C_{\chi }(\widetilde{\mathcal{M}};\Sigma )=C_{\chi }(\mathcal{M};\Sigma ),
\end{equation*}%
i.e. rescaling the observable by nondegenerate $K$ has no effect on the $%
\chi -$capacity of the measurement (\ref{M}) and the optimal ensemble. $%
\square $

The importance of the quantity (\ref{cchi}) is apparent: it is a key for
computing the energy-constrained classical capacity of the channel $\mathcal{%
M}$ which is important quantity characterizing the information-processing
performance of the measurement. Indeed, let
\begin{equation*}
H=a^{\dagger }\epsilon a=\sum_{j,k=1}^{s}\epsilon _{jk}a_{j}^{\dagger }a_{k}
\end{equation*}%
be a quadratic gauge-invariant Hamiltonian, where $\epsilon =\left[ \epsilon
_{jk}\right] $ is positive definite Hermitian matrix, so that the mean
energy of the input d.o. $\rho $ is equal to%
\begin{equation*}
\mathrm{Tr}\rho H=\sum_{j,k=1}^{s}\epsilon _{jk}\Sigma _{kj}=\mathrm{Sp\,}%
\epsilon \Sigma ,
\end{equation*}%
where $\mathrm{Sp}$ denotes trace of $s\times s-$matrices as distinct from
the trace of operators. Then the energy constraint has the form $\mathrm{Sp\,%
}\epsilon \Sigma \leq E,$ where $E$ is a positive number, and the \textit{%
energy-constrained classical capacity} of the channel $\mathcal{M}$ is equal
to
\begin{equation*}
C(\mathcal{M};H,E)=\sup_{\Sigma :\mathrm{Sp\,}\epsilon \Sigma \leq E}C_{\chi
}(\mathcal{M};\Sigma ).
\end{equation*}%
Notice that the additivity issue does not arise here because measurement
channels are entanglement breaking \cite{h1}, \cite{h}. Given an explicit
expression for $C_{\chi }(\mathcal{M};\Sigma )$ such as (\ref{cchig}),
computation of the last supremum is a separate optimization problem which
can be solved analytically in some special cases. For example, if $%
H=\sum_{j}^{s}\hbar \omega _{j}a_{j}^{\dagger }a_{j},$ so that $\epsilon $
is diagonal, and $N=\mathrm{diag}\left[ n_{j}\right] ,$ then the optimal $%
\Sigma $ is also diagonal and its entries $s_{j}$ can be found with a simple
generalization of the \textquotedblleft water-filling
solution\textquotedblright , cf. \cite{cover}, namely%
\begin{equation*}
s_{j}=\left( \nu /\hbar \omega _{j}-n_{j}-1\right) _{+},
\end{equation*}%
where $\nu $ is found from the equation $\sum_{j}^{s}\hbar \omega
_{j}s_{j}=E,$ and%
\begin{equation*}
C(\mathcal{M};H,E)=\sum_{j=1}^{s}\log \left( 1+\frac{s_{j}}{n_{j}+1}\right) .
\end{equation*}

The following result (for observable (\ref{M*})) was conjectured in the
early seventies. In \cite{h6} it was observed that the measurement (\ref{M*}%
) for the Gaussian ensemble (\ref{GE}), (\ref{GE2}) below gives the
information amount (\ref{CX}) which is thus the lower bound for the
accessible information of the ensemble defined as%
\begin{equation*}
A(\mathcal{E}\mathbf{)=}\sup_{\mathcal{M}}I(\mathcal{E},\mathcal{M}),
\end{equation*}%
where the supremum is over all observables $\mathcal{M}$. The conjecture was
that the observable (\ref{M*}) gives the global maximum. In \cite{b1} the
authors verified the necessary local extremality condition for information
based on the first variation derived in \cite{h3}, and in \cite{b2} the
second variation was shown nonpositive\footnote{%
The English versions of these articles were also posted as
arXiv:quant-ph/0511042, arXiv:quant-ph/0511043.}. However to our knowledge
the question of the global maximum was open until now.

\begin{thm}
\label{t2} Let $\mathcal{E}$ \ be the Gaussian ensemble $\left\{ \pi
(d^{2s}z),\rho _{N,z}\right\} ,$ where \footnote{%
For the clarity of proofs we assume that the covariance matrices $\Sigma
,\,N $ are nondegenerate, although this restriction can be relaxed by using
more abstract computations with the quantum characteristic functions.}
\begin{eqnarray}
\pi (d^{2s}z) &=&\exp \left( -z^{\ast }\Sigma ^{-1}z\right) \frac{d^{2s}z}{%
\pi ^{s}\det \Sigma },  \label{GE} \\
\rho _{N,z} &=&D(z)\rho _{N}D(z)^{\dagger }  \label{GE2}
\end{eqnarray}%
is d.o. (\ref{shifted2}) with $\Lambda =N.$

Then the accessible information $A(\mathcal{E})$ of this ensemble is equal
to (\ref{CX}) and is attained on any Gaussian observable of the form%
\begin{equation}
\tilde{M}_{\ast }(d^{2s}z)=D(Kz)\rho _{0}D(Kz)^{\dagger }\frac{\left\vert
\det K\right\vert ^{2}d^{2s}z}{\pi ^{s}},  \label{opk}
\end{equation}%
where $\det K\neq 0,$ in particular, on the observable (\ref{M*}).
\end{thm}

\textit{Proof}. By using (\ref{shifted2}) and convolution of Gaussian
densities, we obtain the average state of the ensemble (\ref{GE}), (\ref{GE2}%
)%
\begin{equation}
\bar{\rho}_{\mathcal{E}}=\int |w\rangle \langle w|\exp \left( -w^{\ast
}\left( \Sigma +N\right) ^{-1}w\right) \frac{d^{2s}w}{\pi ^{s}\det \left(
\Sigma +N\right) }=\rho _{\Sigma +N}.  \label{AV}
\end{equation}%
Computation using (\ref{rchi}) and (\ref{maxh}) gives%
\begin{eqnarray}
I(\mathcal{E},\mathcal{M}_{\ast }) &=&h(p_{\rho _{\Sigma +N}})-h(p_{\rho
_{N}})  \label{LE} \\
&=&\log \det \left( \Sigma +N+I_{s}\right) -\log \det \left( N+I_{s}\right)
\notag \\
&=&\log \det \left( I_{s}+\left( N+I_{s}\right) ^{-1}\Sigma \right)  \notag
\end{eqnarray}%
for the ensemble $\mathcal{E}$ and observable $\mathcal{M}_{\ast }$ defined
by (\ref{M*}), thus giving the lower bound for the accessible information $A(%
\mathcal{E}).$ Any observable (\ref{opk}) gives the same value by (\ref{sca}%
).

We now use the general upper bound 
from the next section:
\begin{equation}
A(\mathcal{E}\mathbf{)}\leq \sup_{\mathcal{E}^{\prime }:\bar{\rho}_{\mathcal{%
E}^{\prime }}=\bar{\rho}_{\mathcal{E}}}I(\mathcal{E}^{\prime },\mathcal{M}%
^{\prime }),  \label{sup1}
\end{equation}%
where $\mathcal{M}^{\prime }$ is observable \emph{dual} to the ensemble $%
\mathcal{E}$ (defined by Eq. (\ref{mprime}) in proposition \ref{p1} below),
and the supremum is taken over all ensembles $\mathcal{E}^{\prime }$
satisfying the condition $\bar{\rho}_{\mathcal{E}^{\prime }}=\bar{\rho}_{%
\mathcal{E}}$. In the case we are considering, this observable will turn out
Gaussian so that we can apply to it theorem \ref{t1} to compute the
right-hand side of the inequality (\ref{sup1}). According to Eq. (\ref%
{mprime}) the dual observable is given by the relation%
\begin{eqnarray}
M^{\prime }(d^{2s}z) &=&\,\bar{\rho}_{\mathcal{E}}^{-1/2}\rho _{N,z}\bar{\rho%
}_{\mathcal{E}}^{-1/2}\pi (d^{2s}z)  \label{MP} \\
&=&\int \,\bar{\rho}_{\mathcal{E}}^{-1/2}|w\rangle \langle w|\bar{\rho}_{%
\mathcal{E}}^{-1/2}\exp \left( -(w-z)^{\ast }N^{-1}(w-z)\right)  \notag \\
&&\times \frac{d^{2s}w}{\pi ^{s}\det N}\pi (d^{2s}z).  \notag
\end{eqnarray}%
By using the decomposition in the normal modes associated with the
orthonormal basis of eigenvectors of the matrix
\begin{equation}
\tilde{\Sigma}=\Sigma +N,  \label{sigtil}
\end{equation}%
we obtain (see (\ref{sqr}) in Appendix A for detail)%
\begin{equation}
\bar{\rho}_{\mathcal{E}}^{-1/2}|w\rangle =\sqrt{\det \left( \tilde{\Sigma}%
+I_{s}\right) }\exp \left\{ \frac{1}{2}w^{\ast }\tilde{\Sigma}^{-1}w\right\}
\left\vert \sqrt{I_{s}+\tilde{\Sigma}^{-1}}w\right\rangle .  \label{VI}
\end{equation}%
Substituting this into (\ref{MP}), we get%
\begin{eqnarray}
&&\int \,\left\vert \sqrt{I_{s}+\tilde{\Sigma}^{-1}}w\right\rangle
\left\langle \sqrt{I_{s}+\tilde{\Sigma}^{-1}}w\right\vert \exp \left[
-w^{\ast }\left( N^{-1}-\tilde{\Sigma}^{-1}\right) w+2\mathrm{Re\,}w^{\ast
}N^{-1}z\right]  \notag \\
&&\times \frac{d^{2s}w\det \left( \tilde{\Sigma}+I_{s}\right) }{\pi ^{s}\det
N}\exp \left( -z^{\ast }N^{-1}z\right) \pi (d^{2s}z).  \label{VII}
\end{eqnarray}%
By making change of variables
\begin{equation*}
w=\sqrt{\left( I_{s}+\tilde{\Sigma}^{-1}\right) ^{-1}}u,\quad \tilde{z}=%
\sqrt{\tilde{\Sigma}\left( \tilde{\Sigma}+I_{s}\right) }\Sigma ^{-1}z=Kz,
\end{equation*}%
and denoting%
\begin{eqnarray}
\tilde{N}^{-1} &=&\sqrt{\left( I_{s}+\tilde{\Sigma}^{-1}\right) ^{-1}}\left[
N^{-1}-\tilde{\Sigma}^{-1}\right] \sqrt{\left( I_{s}+\tilde{\Sigma}%
^{-1}\right) ^{-1}}  \notag \\
&=&\sqrt{\left( I_{s}+\tilde{\Sigma}^{-1}\right) ^{-1}}N^{-1}\Sigma \,\tilde{%
\Sigma}^{-1}\sqrt{\left( I_{s}+\tilde{\Sigma}^{-1}\right) ^{-1}},  \label{N1}
\end{eqnarray}%
we obtain, by arranging the terms in the quadratic form under the exponent
in (\ref{VII}),
\begin{eqnarray*}
M^{\prime }(d^{2s}z) &=&\int |u\rangle \langle u|\exp \left( -(u-\tilde{z}%
)^{\ast }\tilde{N}^{-1}(u-\tilde{z})\right) \frac{d^{2s}u}{\pi ^{s}\det
\tilde{N}}\frac{d^{2s}\tilde{z}}{\pi ^{s}} \\
&=&D(Kz)\rho _{\tilde{N}}D(Kz)^{\dagger }\frac{\left\vert \det K\right\vert
^{2}d^{2s}z}{\pi ^{s}},
\end{eqnarray*}%
which has the same Gaussian form as $\mathcal{\tilde{M}}$ in theorem \ref{t1}%
.

We now compute the supremum in the right-hand side of (\ref{sup1}) by using
theorem \ref{t1} with $N$ replaced by $\tilde{N}$ and $\Sigma $ replaced by $%
\tilde{\Sigma}=\Sigma +N$ from the average state $\bar{\rho}_{\mathcal{E}}$
given by (\ref{AV}). Theorem \ref{t1} then implies%
\begin{equation}
\sup_{\mathcal{E}^{\prime }:\bar{\rho}_{\mathcal{E}^{\prime }}=\bar{\rho}_{%
\mathcal{E}}}I(\mathcal{E}^{\prime },\mathcal{M}^{\prime })\leq C_{\chi }(%
\mathcal{M}^{\prime };\tilde{\Sigma})=\log \det \left( I_{s}+\left( \tilde{N}%
+I_{s}\right) ^{-1}\tilde{\Sigma}\right) .  \label{UE}
\end{equation}%
A computation below shows that
\begin{equation}
\det \left( I_{s}+\left( \tilde{N}+I_{s}\right) ^{-1}\tilde{\Sigma}\right)
=\det \left( I_{s}+T^{-1}\Sigma \left( N+I_{s}\right) ^{-1}T\right) =\det
\left( I_{s}+\left( N+I_{s}\right) ^{-1}\Sigma \right) ,  \label{chu}
\end{equation}%
where $T=\sqrt{\tilde{\Sigma}\left( \tilde{\Sigma}+I_{s}\right) }.$ This
gives the upper estimate for $A(\mathcal{E}\mathbf{)}$ which coincides with
the lower estimate (\ref{LE}), thus proving the theorem.

To prove (\ref{chu}), we obtain from (\ref{N1})%
\begin{equation*}
\tilde{N}+I_{s}=\sqrt{\tilde{\Sigma}\left( \tilde{\Sigma}+I_{s}\right) }%
\left( \Sigma ^{-1}N+\left( \tilde{\Sigma}+I_{s}\right) ^{-1}\right) \sqrt{%
I_{s}+\tilde{\Sigma}^{-1}},
\end{equation*}%
then%
\begin{equation}
\left( \tilde{N}+I_{s}\right) ^{-1}\tilde{\Sigma}=\sqrt{\left( I_{s}+\tilde{%
\Sigma}^{-1}\right) ^{-1}}\left[ \Sigma ^{-1}N+\left( \tilde{\Sigma}%
+I_{s}\right) ^{-1}\right] ^{-1}\sqrt{\left( I_{s}+\tilde{\Sigma}%
^{-1}\right) ^{-1}}.  \label{inter}
\end{equation}%
Substituting%
\begin{equation*}
\Sigma ^{-1}N+\left( \tilde{\Sigma}+I_{s}\right) ^{-1}=\left( \tilde{\Sigma}%
+I_{s}\right) ^{-1}\left( N+I_{s}\right) \Sigma ^{-1}\tilde{\Sigma}
\end{equation*}%
into (\ref{inter}), we obtain%
\begin{equation*}
\left( \tilde{N}+I_{s}\right) ^{-1}\tilde{\Sigma}=T^{-1}\Sigma \left(
N+I_{s}\right) ^{-1}T,
\end{equation*}%
where $T=\sqrt{\tilde{\Sigma}\left( \tilde{\Sigma}+I_{s}\right) },$ hence (%
\ref{chu}) follows. $\square $ 

\section{Ensemble-observable duality}

\label{s3}

Duality between ensembles and observables proved to be an efficient tool in
quantum information theory (see \cite{hall}, \cite{da}, \cite{bl} or \cite%
{h5}). In this section we provide a rigorous infinite-dimensional and
continuous-variables version of this duality used in the proof of theorem %
\ref{t2}.

\begin{prop}
\label{p1} Let $\mathcal{E}=\left\{ \pi (dx),\rho _{x}\right\} $ be an
ensemble and $\mathcal{M}=\left\{ M(dy)\right\} $ an observable such that
\begin{equation}
M(B)=\int_{B}m(y)\mu (dy),  \label{DP}
\end{equation}%
where $\mu (dy)$ is a $\sigma -$finite measure, $m(y)$ is weakly measurable
function with values in the cone of bounded positive operators in $\mathcal{H%
}$ and the integral weakly converges ($B$ is an arbitrary Borel subset of $%
\mathcal{H}$).

Define the dual pair ensemble-observable $(\mathcal{E}^{\prime },\mathcal{M}%
^{\prime })$ by the relations
\begin{equation}
\mathcal{E}^{\prime }:\quad \pi ^{\prime }(B)=\mathop{\rm Tr}\nolimits\bar{%
\rho}_{\mathcal{E}}\,M(B),\quad \rho _{y}^{\prime }=\frac{\bar{\rho}_{%
\mathcal{E}}^{1/2}m(y)\bar{\rho}_{\mathcal{E}}^{1/2}}{\mathop{\rm Tr}%
\nolimits\bar{\rho}_{\mathcal{E}}\,m(y)};  \label{piprime}
\end{equation}%
\begin{equation}
\mathcal{M}^{\prime }:\quad \bra{\psi}M^{\prime }(A)\ket{\psi}=\int_{A}%
\bra{\bar{\rho}_{\mathcal{E}}^{-1/2}\psi}\rho _{x}\ket{\bar{\rho}_{%
\mathcal{E}}^{-1/2}\psi}\pi (dx),  \label{mprime}
\end{equation}%
for $\psi \in \mathrm{ran\,}\bar{\rho}_{\mathcal{E}}^{1/2}\oplus \mathcal{H}%
_{0}$, where $\mathcal{H}_{0}=\mathrm{ker\,}\bar{\rho}_{\mathcal{E}}^{1/2}$
\footnote{%
We use the generalized inverse for $\bar{\rho}_{\mathcal{E}}^{-1/2}$ .}.
Then the average states of both ensembles coincide
\begin{equation}
\bar{\rho}_{\mathcal{E}}=\bar{\rho}_{\mathcal{E}^{\prime }}.  \label{III}
\end{equation}%
Moreover, the joint distribution of $x,y$ is the same for both pairs $(%
\mathcal{E},\mathcal{M})$ and $(\mathcal{E}^{\prime },\mathcal{M}^{\prime })$
so that%
\begin{equation}
I(\mathcal{E},\mathcal{M})=I(\mathcal{E}^{\prime },\mathcal{M}^{\prime }).
\label{II}
\end{equation}
\end{prop}

\textit{Proof}. From (\ref{piprime}) it follows%
\begin{equation*}
\bar{\rho}_{\mathcal{E}^{\prime }}=\int_{\mathcal{Y}}\rho _{y}^{\prime
}\,\pi ^{\prime }(dy)=\int_{\mathcal{Y}}\bar{\rho}_{\mathcal{E}}^{1/2}m(y)%
\bar{\rho}_{\mathcal{E}}^{1/2}\mu (dy)=\bar{\rho}_{\mathcal{E}}^{1/2}\int_{%
\mathcal{Y}}m(y)\mu (dy)\,\bar{\rho}_{\mathcal{E}}^{1/2}=\bar{\rho}_{%
\mathcal{E}}.
\end{equation*}

The definition (\ref{mprime}) implies
\begin{equation*}
0\leq \bra{\psi}M^{\prime }(A)\ket{\psi}\leq \int_{\mathcal{X}}%
\bra{\bar{\rho}_{\mathcal{E}}^{-1/2}\psi}\rho _{x}\ket{\bar{\rho}_{%
\mathcal{E}}^{-1/2}\psi}\pi (dx)=\braket{\psi}{\psi},
\end{equation*}%
for dense domain of $\psi $ implying that $M^{\prime }(A)$ are bounded
positive operators with $M^{\prime }(\mathcal{X})=I$. The definition via
integral also implies $\sigma $-additivity, hence $\mathcal{M}^{\prime }$ is
an observable.

Notice the identity
\begin{equation}
\bar{\rho}_{\mathcal{E}}^{1/2}M^{\prime }(A)\,\bar{\rho}_{\mathcal{E}%
}^{1/2}=\int_{A}\rho _{x}\pi (dx).  \label{ide}
\end{equation}%
Then the joint distribution of $x,y$
\begin{equation*}
P(A\times B)=\int_{A}\pi (dx)\mathop{\rm Tr}\nolimits\rho _{x}M(B)=%
\mathop{\rm Tr}\nolimits\int_{A}\rho _{x}\pi (dx)M(B),
\end{equation*}%
via (\ref{ide}) is equal to
\begin{eqnarray*}
\mathop{\rm Tr}\nolimits\bar{\rho}_{\mathcal{E}}^{1/2}M^{\prime }(A)\,\bar{%
\rho}_{\mathcal{E}}^{1/2}M(B) &=&\mathop{\rm Tr}\nolimits\int_{B}\bar{\rho}_{%
\mathcal{E}}^{1/2}m(y)\bar{\rho}_{\mathcal{E}}^{1/2}\mu (dy)M^{\prime }(A) \\
&=&\int_{B}\pi ^{\prime }(dy)\mathop{\rm Tr}\nolimits\rho _{y}^{\prime
}M^{\prime }(A)=P^{\prime }(A\times B),
\end{eqnarray*}%
hence (\ref{II}) holds. $\square $

The equality (\ref{II}) implies an estimate for the accessible information
of the ensemble $\mathcal{E}$%
\begin{equation*}
A(\mathcal{E}\mathbf{)=}\sup_{\mathcal{M}}I(\mathcal{E},\mathcal{M}),
\end{equation*}%
where the supremum is over all observables $\mathcal{M}$.

\begin{prop}
\label{p2} Let $\mathcal{E}$ \ be a fixed ensemble and $\mathcal{M}^{\prime
} $ be the dual observable, then
\begin{equation}
\sup_{\mathcal{M}}I(\mathcal{E},\mathcal{M})=\sup_{\mathcal{E}^{\prime }:%
\bar{\rho}_{\mathcal{E}^{\prime }}=\bar{\rho}_{\mathcal{E}}}I(\mathcal{E}%
^{\prime },\mathcal{M}^{\prime }),  \label{inf}
\end{equation}%
where the supremum in the right-hand side is taken over all ensembles $%
\mathcal{E}^{\prime }$ satisfying the condition $\bar{\rho}_{\mathcal{E}%
^{\prime }}=\bar{\rho}_{\mathcal{E}}$.
\end{prop}

\textit{Proof}. We first prove the inequality (\ref{sup1}) which was used in
the proof of theorem \ref{t2}. We repeat it here for convenience:
\begin{equation}
A(\mathcal{E}\mathbf{)}\leq \sup_{\mathcal{E}^{\prime }:\bar{\rho}_{\mathcal{%
E}^{\prime }}=\bar{\rho}_{\mathcal{E}}}I(\mathcal{E}^{\prime },\mathcal{M}%
^{\prime }).  \label{sup}
\end{equation}%
For this it is sufficient to show that
\begin{equation}
\sup_{\mathcal{M}}I(\mathcal{E},\mathcal{M})=\sup_{\mathcal{M}:(\ref{DP})}I(%
\mathcal{E},\mathcal{M}),  \label{ss}
\end{equation}%
where on the right the supremum is taken over observables $\mathcal{M}$
which satisfy (\ref{DP}) with respect to some measure $\mu $. Then by using
the proposition \ref{p1} we obtain%
\begin{equation*}
\sup_{\mathcal{M}:(\ref{DP})}I(\mathcal{E},\mathcal{M})=\sup_{\mathcal{E}%
^{\prime }:(\ref{piprime})}I(\mathcal{E}^{\prime },\mathcal{M}^{\prime }),
\end{equation*}%
where in the right-hand side the supremum is taken over ensembles $\mathcal{E%
}^{\prime }$ that can be written in the form $(\ref{piprime})$ for suitable $%
\mathcal{M}$, whence (\ref{sup}) will follow.

Proof of the equality (\ref{ss}) is based on two facts. First, we show that
any observable $\mathcal{M}$ can be approximated by a sequence of
observables $\left\{ \mathcal{M}_{n}\right\} $ satisfying (\ref{DP}) for
some measures $\mu _{n}$. Second, we observe that the information quantity $%
I(\mathcal{E}, \mathcal{M})$ is lower semicontinuous in this approximation.

Let $\mathcal{M}=\{M(dy)\}$ be an observable, and let $\left\{ P_{n}\right\}
$ be a nondecreasing sequence of projections in $\mathcal{H}$ such that $%
P_{n}\uparrow I$ as $n\rightarrow \infty .$ Define the measure $\mu _{n}(B)=%
\mathrm{Tr}P_{n}M(B)$ and the sequence of observables
\begin{equation}
M_{n}(B)=P_{n}M(B)P_{n}\oplus \left( I-P_{n}\right) \mu _{n}(B)/\mathrm{Tr}%
\,P_{n}.  \label{Mn}
\end{equation}%
Then $\mathcal{M}_{n}=\{M_{n}(dy)\}$ satisfies (\ref{DP}) with the measure $%
\mu _{n}(B).$ Indeed, $0\leq P_{n}M(B)P_{n}\leq P_{n}\mu _{n}(B)$ and the
second term in the direct sum (\ref{Mn}) is dominated by $\left(
I-P_{n}\right) \mu _{n}(B).$ Hence, by an operator version of Radon-Nikodym
theorem, $M_{n}(B)=\int_{B}m_{n}(y)\mu _{n}(dy),$ with $\left\Vert
m_{n}(y)\right\Vert \leq 1$ (\textrm{mod} $\mu _{n}$).

For arbitrary d.o. $\rho $ and arbitrary Borel $B\subseteq \mathcal{Y}$
\begin{eqnarray}
&&\left\vert \mathrm{Tr}\,\rho \,M(B)-\mathrm{Tr}\,\rho \,M_{n}(B)\right\vert
\label{app} \\
&\leq &\left\vert \mathrm{Tr}\,\rho \,\left( M(B)-P_{n}M(B)P_{n}\right)
\right\vert +\,\left( \,\mu _{n}(B)/\mathrm{Tr}\,P_{n}\right) \mathrm{Tr}%
\,\left( I-P_{n}\right) \rho  \notag \\
&\leq &\left\vert \mathrm{Tr}\,\rho \left( I-P_{n}\right) \,M(B)\right\vert
+\left\vert \mathrm{Tr}\,\rho \,P_{n}M(B)\left( I-P_{n}\right) \right\vert
+\,\,\mathrm{Tr}\,\left( I-P_{n}\right) \rho  \notag \\
&\leq &3\left\Vert \left( I-P_{n}\right) \rho \right\Vert _{1}\rightarrow 0
\notag
\end{eqnarray}%
as $n\rightarrow \infty .$

Now let $\mathcal{V=}\left\{ B_{k}\right\} $ be a finite decomposition of
the space $\mathcal{Y}$ into Borel subsets $B_{k}.$ Define the finitely
valued \textquotedblleft coarse-grained\textquotedblright\ observable $%
\mathcal{M}_{\mathcal{V}}=\left\{ M(B_{k})\right\} .$ A general result of
classical information theory (cf. \cite{dobr}) implies%
\begin{equation*}
I(\mathcal{E},\mathcal{M})=\sup_{\mathcal{V}}I(\mathcal{E},\mathcal{M}_{%
\mathcal{V}}),
\end{equation*}%
where the supremum is taken over all the decompositions $\mathcal{V}$. We
will prove that for a fixed $\mathcal{V}$ the functional $I(\mathcal{E},%
\mathcal{M}_{\mathcal{V}})$ is continuous with respect to the approximation (%
\ref{app}), then it will follow that $I(\mathcal{E},\mathcal{M})$ is lower
semicontinuous. Denoting $P_{\rho }(B)=\mathrm{Tr}\,\rho \,M(B),$ we have%
\begin{equation*}
I(\mathcal{E},\mathcal{M}_{\mathcal{V}})=-\sum_{k}P_{\bar{\rho}_{\mathcal{E}%
}}(B_{k})\log P_{\bar{\rho}_{\mathcal{E}}}(B_{k})+\int \pi
(dx)\sum_{k}P_{\rho _{x}}(B_{k})\log P_{\rho _{x}}(B_{k}).
\end{equation*}%
When we approximate $\mathcal{M}$ by $\mathcal{M}_{n},$ the first term
converges by (\ref{app}) and by continuity of the Shannon entropy. In the
second term the integrand converges pointwise by (\ref{app}) and it is
uniformly bounded because $-e^{-1}\log e\leq P\log P\leq 0$ for $0\leq P\leq
1.$ This finishes the proof of (\ref{ss}) and hence of (\ref{sup}).

Let us now prove the stronger result: the equality (\ref{inf}), by showing
that any ensemble $\mathcal{E}^{\prime}=\left\{ \pi (dy),\rho _{y}\right\} $
with fixed average state $\bar{\rho}_{\mathcal{E}^{\prime}}=\bar{\rho}_{%
\mathcal{E}}$ can be approximated by ensembles of the form (\ref{piprime}).
First, if $\bar{\rho}_{\mathcal{E}}$ has finite rank, the problem reduces to
finite dimensional one which is easily solved. Therefore assume that the
rank of $\bar{\rho}_{\mathcal{E}}$ is infinite (for simplicity we can assume
that $\bar{\rho}_{\mathcal{E}}$ is nondegenerated). Let $P_{n}$ be the
projection onto the eigenspace of $\bar{\rho}_{\mathcal{E}}$ corresponding
to $n$ largest eigenvalues. Let
\begin{equation*}
m_{n}(y)=P_{n}\bar{\rho}_{\mathcal{E}}^{-1/2}\rho _{y}\bar{\rho}_{\mathcal{E}%
}^{-1/2}P_{n}\oplus \left( I-P_{n}\right) \frac{1-\mathrm{Tr}\,\rho _{y}P_{n}%
}{1-\mathrm{Tr}\,\bar{\rho}_{\mathcal{E}}P_{n}},
\end{equation*}%
then%
\begin{equation*}
0\leq m_{n}(y)\leq \lambda _{n}^{-1}P_{n}\oplus \left( I-P_{n}\right) /%
\mathrm{Tr}\,\bar{\rho}_{\mathcal{E}}\left( I-P_{n}\right) ,
\end{equation*}%
where $\lambda _{n}$ is the smallest eigenvalue for eigenvectors in the
range of $P_{n}.$ Then%
\begin{equation}
\int m_{n}(y)\pi (dy)=P_{n}\oplus \left( I-P_{n}\right) =I,  \label{MI}
\end{equation}%
hence $M(B)=\int_{B}m_{n}(y)\,\pi (dy)$ is an observable. Moreover, $\mathrm{%
Tr\,}\bar{\rho}_{\mathcal{E}}\,m_{n}(y)=1$ (\textrm{mod} $\pi $). Define
ensemble $\mathcal{E}_{n}^{\prime }=\left\{ \pi _{n}(dy),\rho
_{y}^{n}\right\} $ by taking $\pi _{n}(dy)=\pi (dy)$ and%
\begin{equation*}
\rho _{y}^{n}=\bar{\rho}_{\mathcal{E}}^{1/2}m_{n}(y)\bar{\rho}_{\mathcal{E}%
}^{1/2},
\end{equation*}%
then by (\ref{MI}) the average state of $\mathcal{E}_{n}^{\prime }$ is $\bar{%
\rho}_{\mathcal{E}}.$ Ensemble $\mathcal{E}_{n}^{\prime }$ has the required
form (\ref{piprime}). Moreover, for any observable $\mathcal{M}^{\prime
}=\left\{ M^{\prime}(dx)\right\} $ the joint probability%
\begin{equation*}
P_{n}^{\prime }(A\times B)=\int_{B}\,\pi (dy)\mathrm{Tr}\,\rho
_{y}^{n}M^{\prime}(A)\rightarrow \int_{B}\,\pi (dy)\mathrm{Tr}\,\rho
_{y}M^{\prime}(A)=P(A\times B).
\end{equation*}%
Indeed,
\begin{equation*}
\mathrm{Tr}\,\rho _{y}^{n}M^{\prime}(A)=\mathrm{Tr}\,P_{n}\,\rho
_{y}\,P_{n}M^{\prime}(A)+\mathrm{Tr}\,\bar{\rho}_{\mathcal{E}}\left(
I-P_{n}\right) M^{\prime}(A)\frac{1-\mathrm{Tr}\,\rho _{y}P_{n}}{1-\mathrm{Tr%
}\,\bar{\rho}_{\mathcal{E}}P_{n}}\rightarrow \mathrm{Tr}\,\rho
_{y}M^{\prime}(A)
\end{equation*}%
pointwise, remaining uniformly bounded by 1. For any finite decomposition $%
\mathcal{V=}\left\{ A_{k}\right\} $ of the space $\mathcal{X}$ and $\mathcal{%
V^{\prime }=}\left\{ B_{k}\right\} $ of the space $\mathcal{Y}$, the
\textquotedblleft coarse-grained\textquotedblright\ mutual information is
continuous and the mutual information is lower semicontinuous by the
argument in the proof above, hence
\begin{equation*}
\lim \inf_{n\rightarrow \infty }\,I(\mathcal{E}_{n}^{\prime },\mathcal{M}%
^{\prime })\geq I(\mathcal{E}^{\prime },\mathcal{M}^{\prime }).
\end{equation*}%
It implies finally the equality (\ref{inf}). $\square $

\section{Conclusion}

We have considered quantum Gaussian multimode system with the global gauge
symmetry and obtained explicit formula for the classical capacity of a
Gaussian observable which describes statistics of a noisy heterodyne
measurement in such a system. We have shown that the capacity is attained by
a Gaussian ensemble of coherent states. The condition of gauge covariance
was relaxed in our recent paper \cite{hk}, where the formula was generalized
to the case where only certain \textquotedblleft threshold
condition\textquotedblright\ is fulfilled. Our second result gives explicit
expression for the accessible information of a gauge-invariant Gaussian
ensemble, and shows that it is attained by the multimode generalization of
the (ideal) heterodyne measurement, solving a conjecture going back to the
seventies. Moreover, the same value is attained by any multimode scaling of
the measurement, illustrating the high degeneracy of the maximum
characteristic to such kind of \textquotedblleft quantum Gaussian
optimizer\textquotedblright\ problems. A natural question of extensions of
this result to quantum Gaussian systems without gauge symmetry, or even
without any \textquotedblleft threshold condition\textquotedblright\ remains
open for investigation.

\section{Appendix A}

Let $\rho ,\sigma $ be two d.o., then, generalizing (\ref{pro}), the relation%
\begin{equation*}
\mathrm{Tr}\,\rho \,D(z)\,\sigma D(z)^{\dagger }
\end{equation*}%
defines a p.d. on $\mathbb{C}^{s}.$ Its classical characteristic function
expressed via the symplectic Fourier transform is
\begin{eqnarray}
&&\int \exp \left[ 2i\mathrm{Im\,}z^{\ast }w\right] \,\mathrm{Tr}\,\rho
\,D(z)\,\sigma D(z)^{\dagger }\frac{d^{2s}z}{\pi ^{s}}  \notag \\
&=&\int \exp \left[ 2i\mathrm{Im\,}z^{\ast }w\right] \,\int \mathrm{Tr}%
\,\rho \,D(u)\overline{\exp \left[ 2i\mathrm{Im\,}z^{\ast }u\right] \,%
\mathrm{Tr\,}\sigma D(u)}  \label{cf2} \\
&=&\int \int \exp \left[ 2i\mathrm{Im\,}z^{\ast }(w-u)\right] \frac{d^{2s}z}{%
\pi ^{2s}}\mathrm{Tr}\,\rho \,D(u)\mathrm{Tr}\,\sigma ^{\top }D(\bar{u}%
)d^{2s}u  \label{cf3} \\
&=&\mathrm{Tr}\,\rho \,D(w)\mathrm{Tr}\,\sigma ^{\top }D(\bar{w}),
\label{cf}
\end{eqnarray}%
where in (\ref{cf2}) we used (\ref{ccr}) and the Parceval identity (\ref%
{parc}), and in (\ref{cf3}) the transposition $\sigma \rightarrow \sigma
^{\top }$ is defined by the relation
\begin{equation}
\mathrm{Tr}\,\sigma ^{\top }\,D(\bar{w})=\overline{\mathrm{Tr}\,\sigma D(w)}%
,\quad w\in \mathbb{C}^{s}.  \label{trans}
\end{equation}%
The expression (\ref{cf}) can be rewritten as%
\begin{eqnarray}
&&\mathrm{Tr}\,\left( \rho \otimes \,\sigma ^{\top }\right) \left(
D(w)\otimes D(\bar{w})\right)  \notag \\
&=&\mathrm{Tr}\,\left( \rho \otimes \,\sigma ^{\top }\right) \exp \left(
\alpha ^{\dagger }w-w^{\ast }\alpha \right) ,  \label{ccf}
\end{eqnarray}%
where $\alpha ,\,\alpha ^{\dagger }$ act in $\mathcal{H\otimes H}_{0},$ $%
\mathcal{H}_{0}\simeq \mathcal{H}$ is the Hilbert space of the ancillary
system$.$ The vectors $\alpha ,\,\alpha ^{\dagger }$ have the components

\begin{eqnarray*}
\alpha _{j} &=&a_{j}\otimes I_{0}+I\otimes a_{0j}^{\dagger }, \\
\alpha _{k}^{\dagger } &=&a_{k}^{\dagger }\otimes I_{0}+I\otimes a_{0k},
\end{eqnarray*}%
where $a_{0j}^{{}},\,a_{0j}^{\dagger }$ are annihilation-creation \
operators in $\mathcal{H}_{0}.$ These components are commuting normal
operators, so that they have joint probability distribution with the
classical characteristic function (\ref{ccf}). Assuming that $\rho \in
\mathfrak{S}(\Sigma ),\,\sigma \in \mathfrak{S}(N),$ let us find the complex
covariance matrix of this distribution. It has the components%
\begin{eqnarray*}
\mathsf{M}\,\alpha _{j}\,\alpha _{k}^{\dagger } &=&\mathrm{Tr}\,\alpha
_{j}\,\left( \rho \otimes \,\sigma ^{\top }\right) \alpha _{k}^{\dagger } \\
&=&\mathrm{Tr}\,\mathsf{\,}a_{j}\rho \,a_{k}^{\dagger }\,+\mathrm{Tr}\mathsf{%
\,}a_{0j}^{\dagger }\,\sigma ^{\top }a_{0k} \\
&=&\mathrm{Tr}\,\mathsf{\,}a_{j}\rho \,a_{k}^{\dagger }\,+\mathrm{Tr}%
\,\sigma ^{\top }a_{0k}a_{0j}^{\dagger } \\
&=&\mathrm{Tr}\,\mathsf{\,}a_{j}\rho \,a_{k}^{\dagger }\,+\mathrm{Tr}%
\,\sigma ^{\top }\left( a_{0j}^{\dagger }a_{0k}+\delta _{jk}I_{0}\right) \\
&=&\mathrm{Tr}\,\mathsf{\,}a_{j}\rho \,a_{k}^{\dagger }+\mathrm{Tr}%
\,a_{0k}\sigma ^{\top }a_{0j}^{\dagger }+\delta _{jk} \\
&=&\mathrm{Tr}\,\mathsf{\,}a_{j}\rho \,a_{k}^{\dagger }+\mathrm{Tr}\,\mathsf{%
\,}a_{0j}\sigma \,a_{0k}^{\dagger }+\delta _{jk}.
\end{eqnarray*}%
Thus the complex covariance matrix is
\begin{equation*}
\mathsf{M}\,\alpha \,\alpha ^{\dagger }=\Sigma +N+I_{s}=\tilde{\Sigma}+I_{s}.
\end{equation*}

Let $\left\{ e_{k}\right\} $ be an orthonormal basis in $\mathbb{C}^{s},$
and let$\ z=\sum_{k=1}^{s}\zeta _{k}e_{k}$ be a decomposition of the vector $%
z$ in this basis. Then $a^{\dagger }=\sum_{k=1}^{s}\zeta _{k}b_{k}^{\dagger
},$ where $a^{\dagger }e_{k}=b_{k}^{\dagger }$ are the new creation
operators, corresponding to the modes associated with the basis $\left\{
e_{k}\right\} .$ Let $|n_{k}\rangle $ be the eigenvector of the $k-$th mode
number operator $b_{k}^{\dagger }b_{k}$, corresponding to the eigenvalue $%
n_{k}(=0,1,\dots ).$ Then one has tensor product decomposition of a coherent
state vector
\begin{equation}
|z\rangle =\otimes _{k=1}^{s}\sum_{n_{k}=0}^{\infty }\frac{\zeta _{k}^{n_{k}}%
}{\sqrt{n_{k}!}}\,|n_{k}\rangle \exp \left( -\frac{|\zeta _{k}|^{2}}{2}%
\right) .  \label{tenscs}
\end{equation}%
If $\left\{ e_{k}\right\} $ is the basis of eigenvectors of the covariance
matrix $\Lambda $ of the Gaussian d.o. $\rho _{\Lambda },$ with the
corresponding eigenvalues $\lambda _{k},$ then%
\begin{equation}
\rho _{\Lambda }=\otimes _{k=1}^{s}\frac{1}{\lambda _{k}+1}%
\sum_{n_{k}=0}^{\infty }\left( \frac{\lambda _{k}}{\lambda _{k}+1}\right)
^{n_{k}}|n_{k}\rangle \langle n_{k}|.  \label{tensdo}
\end{equation}%
It follows that $\rho _{\Lambda }^{-1/2}|z\rangle $ is given by the
expression
\begin{equation*}
\otimes _{k=1}^{s}\sqrt{\lambda _{k}+1}\sum_{n_{k}=0}^{\infty }\frac{\left(
\sqrt{1+\lambda _{k}^{-1}}\zeta _{k}\right) ^{n_{k}}}{\sqrt{n_{k}!}}%
\,|n_{k}\rangle \exp \left( -\frac{(1+\lambda _{k}^{-1})|\zeta _{k}|^{2}}{2}+%
\frac{|\zeta _{k}|^{2}}{2\lambda _{k}}\right)
\end{equation*}%
\begin{equation}
=\sqrt{\Lambda +I_{s}}\exp \left\{ \frac{1}{2}z^{\ast }\Lambda
^{-1}z\right\} \left\vert \sqrt{I_{s}+\Lambda ^{-1}}z\right\rangle .
\label{sqr}
\end{equation}%
The formula (\ref{VI}) is obtained by choosing the basis of eigenvectors of
the covariance matrix $\Lambda =\tilde{\Sigma}$ and then using this
expression.

\textbf{Acknowledgment}. The work was supported by the grant of Russian
Scientific Foundation (project No 19-11-00086). The author is grateful to
M.E. Shirokov, G.G. Amosov, S.N. Filippov and anonymous referees for useful
remarks.

\end{document}